\documentstyle[12pt]{article}
\begin{document}

\begin{center}
{\large\bf Structure of Low-Energy Effective Action in N=4
Supersymmetric Yang-Mills Theories}

\vspace{0.5cm}
I.L. Buchbinder

\vspace{0.3cm}
Department of Theoretical Physics

Tomsk State Pedagogical University

Tomsk, 634041, Russia
\end{center}

\vspace{0.8cm}
\begin{abstract}
We study a problem of low-energy effective action in N=4 super
Yang-Mills theories. Using harmonic superspace approach we consider N=4
SYM in terms of unconstrained N=2 superfield and apply N=2 background
field method to finding effective action for N=4 SU(n) SYM broken down to
U(n)$^{n-1}$. General structure of leading low-energy corrections to
effective action is discussed and calculational procedure for their
explicit finding is presented.
\end{abstract}

\vspace{0.8cm}
\begin{center}
{\large\bf 1. Introduction}
\end{center}

Low-energy structure of quantum supersymmetric field theories is
described by the effective lagrangians of two types: chiral and general
or holomorphic and non-holomorphic. Non-holomorphic or general
contributions to effective action are given by integrals over full
superspace while holomorphic or chiral contributions are given by
integrals over chiral subspace of superspace. As a result, the
effective action in supersymmetric theories should have the form
$$
\Gamma=\left(\int\limits_{\mbox{\tiny chiral\ subspace}}{\cal F} + c.c.
\right)+
\int\limits_{\mbox{\tiny full\ superspace}}{\cal H}+\dots\eqno(1.1)
$$
where the dots mean the terms in effective action depending on
covariant derivatives of the superfields. The complex chiral superfield
${\cal F}$ is called holomorphic or chiral effective potential and real
superfield \ ${\cal H}$ \ is called non-holomorphic or general effective
potential. Thus, the notions of holomorphic and non-holomorphic
effective potentials are very generic and characterizing in principle
any superfield model. We point out that a possibility of holomorphic
corrections to effective action was firstly demonstrated in refs [1-3]
( see also [4,5]) for N=1 SUSY and in refs [6,7] for N=2 SUSY.

The modern interest to structure of low-energy effective action in
extended supersymmetric theories was inspired by the seminal papers [8]
where exact instanton contribution to holomorphic effective potential
has been found for N=2 SU(2) super Yang-Mills theory. The models with
gauge groups SU(n) and SO(n) were considered in refs [9]. One can show
that in generic N=2 SUSY models namely the holomorphic effective
potential is leading low-energy contribution. Non-holomorphic potential
is next to leading correction. A detailed investigation of structure of
low-energy effective action for various N=2 SUSY theories has been
undertaken in refs [10-17].

A further study of quantum aspects of supersymmetric field models leads
to problem of effective action in N=4 SUSY theories. These theories
being maximally extended global supersymmetric models posses the
remarkable properties on quantum level. We list only two of them:

\begin{itemize}
\item[(i)]
N=4 super Yang-Mills model is finite quantum field theory,
\item[(ii)] N=4
super Yang-Mills model is superconformal invariant theory and hence,
its effective action can not depend on any scale. These properties
allow to analyze a general form of low-energy effective action and see
that it changes drastically in compare with generic N=2 super
Yang-Mills theories.
\end{itemize}

Analysis of structure of low-energy effective action in N=4 SU(2) SYM
model spontaneously broken down to U(1) has been fulfilled in recent
paper by Dine and Seiberg [18] (see also [26] for the other gauge
groups). They have investigated a part of effective action depending on
N=2 superfield strengths $W$, $\bar W$ and shown

\begin{itemize}
\item[(i)]
Holomorphic quantum corrections vanish identically
in N=4 SYM. Therefore, namely non-holomorphic effective potential is
leading low-energy contribution to effective action.

\item[(ii)] Non-holomorphic effective potential ${\cal H}(W,\bar W)$
can be found on the base of the properties of quantum N=4 SYM theory up
to a coefficient. All perturbative or non-perturbative corrections do
not influence on functional form of ${\cal H}(W,\bar W)$ and concern
only this coefficient.
\end{itemize}

The approaches to direct calculation of non-holomorphic effective
potential including the above coefficient have been developed in refs
[18-21], extensions for gauge group SU(n) spontaneously broken to
maximal torus have been given in refs [23-25] ( see also [22] where
some bosonic contributions to low-energy effective action have been
found).

This paper is a brief review of our approach [21,24] to derivation of
non-holomorphic effective potential in N=4 SYM theories.


\begin{center}
{\large\bf 2. N=4 super Yang-Mills theory in terms of
unconstrained harmonic superfields}
\end{center}

As well known, the most powerful and adequate approach to investigate
the quantum aspects of supersymmetric field theories is formulation of
these theories in terms of unconstrained superfields carrying out a
representation of supersymmetry. Unfortunately such a manifestly N=4
supersymmetric formulation for N=4 Yang-Mills theory is still unknown.
A purpose of this paper is study a structure of low-energy effective
action for N=4 SYM as a functional of N=2 superfield strengths. In
this case it is sufficient to realize the N=4 SYM theory as a theory
of N=2 unconstrained superfields.
It is naturally achieved within harmonic superspace.
The N=2 harmonic superspace [28] is
the only manifestly N=2 supersymmetric formalism allowing to describe
general N=2 supersymmetric field theories in terms of unconstrained N=2
superfields.  This approach has been successfully applied to problem of
effective action in various N=2 models in recent works [12, 13, 15-17,
21, 24]. We discuss here the results of the papers [21, 24].{}

{}From point of view of N=2 SUSY, the N=4 Yang-Mills theory describes
interaction of N=2 vector multiplet with hypermultiplet in adjoint
representation. Within harmonic superspace approach, the vector
multiplet is realized by unconstrained analytic gauge superfield
$V^{++}$. As to hypermultiplet, it can be described either by areal
unconstrained superfield $\omega$ ($\omega$-hypermultiplet) or by a
complex unconstrained analytic superfield $q^+$ and its conjugate
($q$-hypermultiplet).

In the $\omega$-hypermultiplet realization, the classical action of N=4
SYM model has the form
$$
S[V^{++},\omega]=\frac{1}{2g^2}{\rm tr}\int d^4xd^4\theta W^2-
\frac{1}{2g^2}{\rm tr}\int d\zeta^{(-4)}\nabla^{++}\omega
\nabla^{++}\omega \eqno(2.1)
$$
The first terms here is pure N=2 SYM action and the second term is
action $\omega$-hypermultiplet. In $q$-hypermultiplet realization, the
action of the N=4 SYM model looks like this
$$
S[V^{++},q^+,\stackrel{\smile}{q}^+]=
\frac{1}{2g^2}{\rm tr}\int d^4xd^4\theta W^2-
\frac{1}{2g^2}{\rm tr}\int d\zeta^{(-4)}q^{+i}\nabla^{++}q^+_i
\eqno(2.2)
$$
where
$$
q^+_i=(q^+,\stackrel{\smile}{q}^+),\qquad
q^{i+}=\varepsilon^{ij}q^+_j=(\stackrel{\smile}{q}^+,-q^+)
\eqno(2.3)
$$
All other denotions are given in ref [28].

Both models (2.1, 2.2) are classically equivalent and manifestly N=2
supersymmetric by construction. However, as has been shown in refs
[28], both these models posses hidden N=2 supersymmetry and as a result
they actually are N=4 supersymmetric.

\begin{center}
{\large\bf 3. General form of non-holomorphic effective
potential.}
\end{center}

We study the effective action $\Gamma$ (1.1) for N=4 SYM with gauge
group SU(2) spontaneously broken down to U(1). This effective action is
considered as a functional of N=2 superfield strengths $W$ and $\bar W$.
Then holomorphic effective potential ${\cal F}$ depends on chiral
superfield $W$ and it is integrated over chiral subspace of N=2
superspace with the measure $d^4x\,d^4\theta$. Non-holomorphic
effective potential ${\cal H}$ depends on both $W$ and $\bar W$. It is
integrated over full N=2 superspace with the measure $d^4x\,d^8\theta$.

Let us begin with dimensional
analysis of low-energy effective action. Taking
into account the mass dimensions of $W$, ${\cal F}(W)$, ${\cal
H}(W,\bar W)$ and the superspace measures $d^4x\,d^4\theta$ and
$d^4x\,d^8\theta$ ones write
$$
{\cal F}(W)=W^2f\left(\frac{W}{\Lambda}\right),\qquad
{\cal H}(W,\bar W)={\cal H}\left(\frac{W}{\Lambda},
\frac{\bar W}{\Lambda}\right) \eqno(3.1)
$$
where $\Lambda$ is some scale and $f(\frac{W}{\Lambda})$ and
${\cal H}(\frac{W}{\Lambda},\frac{\bar W}{\Lambda})$ are the
dimensionless functions of their arguments.

Due to remarkable properties of N=4 SYM in quantum domain, the
effective action is scale independent. Therefore
$$
\Lambda\frac{d}{d\Lambda}
\int d^4x\,d^4\theta W^2f\left(\frac{W}{\Lambda}\right)=0
\eqno(3.2)
$$
$$
\Lambda\frac{d}{d\Lambda}
\int d^4x\,d^8\theta{\cal H}\left(\frac{W}{\Lambda},
\frac{\bar W}{\Lambda}\right)=0 \eqno(3.3)
$$
Eq. (3.2) leads to $f(\frac{W}{\Lambda})=const$. Eq (3.3) reads
$$
\Lambda\frac{d}{d\Lambda}{\cal H}=g\left(\frac{W}{\Lambda}\right)+
\bar g\left(\frac{\bar W}{\Lambda}\right)\eqno (3.4)
$$
Here $g$ is arbitrary chiral function of chiral superfield
$\frac{W}{\Lambda}$ and $\bar g$ is conjugate function. The integral of
$g$ and $\bar g$ over full N=2 superspace vanishes and eq (3.3) takes
place for any $g$ and $\bar g$.

Since $f(\frac{W}{\Lambda})=const$ the holomorphic effective potential
${\cal F}(W)$ has classical form $W^2$. General solution of Eq (3.4) is
written as follows
$$
{\cal H}\left(\frac{W}{\Lambda},\frac{\bar W}{\Lambda}\right)=
c\log\frac{W^2}{\Lambda^2}\log\frac{\bar W^2}{\Lambda^2}\eqno(3.5)
$$
with arbitrary coefficient $c$. As a result, holomorphic effective
potential is trivial in N=4 SYM theory. Therefore, namely
non-holomorphic effective potential is leading low-energy quantum
contribution to effective action.  Moreover, the non-holomorphic
effective potential is found exactly up to coefficient and given by
eq. (3.5) [18]. Any perturbative or non-perturbative quantum
corrections are included into a single constant $c$.

However, this result immediately face the problems:

\begin{itemize}
\item[1.] Is there exist a calculational procedure allowing to derive
${\cal H}(W/\Lambda,\bar W/\Lambda)$ in form (3.5) within a model?

\item[2.] What is value of $c$? If $c=0$, the non-holomorphic effective
potential
vanishes and low-energy effective action in N=4 SYM is defined the next
terms in expansion of effective action (1.1) in derivatives.

\item[3.] What is structure of non-holomorphic effective potential for
the other then SU(2) gauge groups?
\end{itemize}

The answers all these questions have been given in refs [19-25].
Further we are going to discuss a general manifestly N=2 supersymmetric
and gauge invariant procedure of deriving the non-holomorphic
effective potential in one-loop approximation [21,24]. This procedure
is based on the following points

\begin{itemize}
\item[(1).] Formulation of N=4 SYM theory in terms of N=2 unconstrained
superfields in harmonic superspace [28].

\item[(2).] N=2 background field method [15,17] providing manifest gauge
invariance on all steps of calculations.

\item[(3).] Identical transformation of path integral for effective
action over N=2 superfields to path integral over some N=1 superfields.
This point is nothing more then replacement of variables in path
integral.

\item[(4).] Superfield proper-time technique [4] which is manifestly
covariant method for calculating effective action in superfield
theories.
\end{itemize}

Next section is devoted to some details of calculating non-holomorphic
effective potential.

\begin{center}
{\large\bf 4. Calculaltion of non-holomorphic effective potential}
\end{center}

We study effective action for the classically equivalent theories (2.1,
2.2) within N=2 background field method [15-17]. We assume also that
the gauge group of these theories is SU(n). In accordance with
background field method [15-17], the one-loop effective action in both
realizations of N=4 SYM is given by
$$
\Gamma^{(1)}[V^{++}]=\frac{i}{2}{\rm Tr}_{(2,2)}
\log\stackrel{\frown}{\Box}-\frac{i}{2}{\rm Tr}_{(4,0)}
\log\stackrel{\frown}{\Box} \eqno(4.1)
$$
where $\stackrel{\frown}{\Box}$ is the analytic d'Alambertian
introduced in ref [15].
$$
\begin{array}{rcl}
\stackrel{\frown}{\Box}&=&{\cal D}^m{\cal D}_m+
\displaystyle\frac{i}{2}({\cal D}^{+\alpha}W){\cal D}^-_\alpha+
\displaystyle\frac{i}{2}(\bar{\cal D}^+_{\dot{\alpha}}\bar W){\bar
{\cal D}}^{-\dot{\alpha}}-\\
&-&\displaystyle\frac{i}{4}({\cal D}^{+\alpha}
{\cal D}^+_\alpha W){\cal D}^{--}+
\displaystyle\frac{i}{8}[{\cal D}^{+\alpha},{\cal
D^-_\alpha}]W+\displaystyle\frac{i}{2} \{\bar{W},W\}
\end{array}\eqno(4.2)
$$
The formal definitions of the
${\rm Tr}_{(2,2)}\log\stackrel{\frown}{\nabla}$ and
${\rm Tr}_{(4,0)}\log\stackrel{\frown}{\nabla}$ are given in ref [21].

Our purpose is finding of non-holomorphic effective potential ${\cal
H}(W,\bar W)$ where the constant superfields $W$ and $\bar W$ belong to
Cartan subalgebra of the gauge group SU(n). Therefore, for calculation
of ${\cal H}(W,\bar W)$ it is sufficient to consider on-shell
background
$$
{\cal D}^{\alpha(i}{\cal D}^{j)}_\alpha W=0 \eqno(4.3)
$$
In this case the one-loop effective action (4.1) can be written in the
form [21]
$$
\exp(i\Gamma^{(1)})=\frac{\int{\cal D}{\cal F}^{++}\exp\left\{
-\frac{i}{2}{\rm tr}\int d\zeta^{(-4)}{\cal F}^{++}\stackrel{\frown}
{\Box}{\cal F}^{(++)}\right\}}
{\int {\cal D}{\cal F}^{++}\exp\left\{-\frac{i}{2}{\rm tr}\int
d\zeta^{(-4)}{\cal F}^{++}{\cal F}^{++}\right\}}\eqno(4.4)
$$
The superfield ${\cal F}^{++}$ belonging to the adjoint representation
looks like ${\cal F}^{++} = {\cal F}^{ij}u^+_iu^+_j$ with
$u^+_i$ be the harmonics [28] and ${\cal F}^{ij}={\cal F}^{ji}$
satisfying the constraints
$$
{\cal D}^{(i}_\alpha{\cal F}^{jk)}=\bar{\cal D}^{(i}_{\dot{\alpha}}
{\cal F}^{jk)}=0,\qquad
\bar{\cal F}^{ij}={\cal F}_{ij}\eqno(4.5)
$$

The next step is transformation of the path integral (4.4) to one over
unconstrained N=1 superfields. This point is treated as replacement of
variables in path integral (4.4). We introduce the N=1 projections of
$W$ ( see the details in refs [21, 24]). As a result one obtains
$$
\Gamma^{(1)}=\sum\limits_{k<l}\Gamma_{kl},\qquad
\Gamma_{kl}=i{\rm Tr}\log\Delta_{kl} \eqno(4.6)
$$
where
$$
\Delta_{kl}={\cal D}^m{\cal D}_m-(W^{k\alpha}-W^{l\alpha})
{\cal D}_\alpha+(\bar{W}^k_{\dot{\alpha}}-\bar{W}^l_{\dot{\alpha}})
\bar{\cal D}^{\dot{\alpha}}+|\Phi^k-\Phi^l|^2 \eqno(4.7)
$$
and ${\cal D}_m$, ${\cal D}_\alpha$, $\bar{\cal D}_{\dot\alpha}$ are
the supercovariant derivatives. Here
$$
\Phi={\rm diag}(\Phi^1,\Phi^2,\dots,\Phi^n),\qquad
\sum\limits_{k=1}^n \Phi^k=0. \eqno(4.8)
$$
$$
W_\alpha={\rm diag}(W_\alpha^1,\dots,W_\alpha^n),\qquad
\sum\limits_{k=1}^n W_\alpha^k = 0
$$
The operator (4.7) has been introduced in ref [24].

Thus, we get a problem of effective action associated with N=1 operator
(4.7). Such a problem can be investigated within N=1 superfield
proper-time technique [4]. Application of this technique leads to
lowest contribution to effective action in the form
$$
\Gamma_{kl}=\frac{1}{(4\pi)^2}\int d^8z\frac{W^{\alpha kl}W^{kl}_\alpha
\bar{W}^{kl}_{\dot{\alpha}}\bar{W}^{\dot{\alpha}kl}}
{(\Phi^{kl})^2(\bar{\Phi}^{kl})^2} \eqno(4.9)
$$
where
$$
\Phi^{kl}=\Phi^k-\Phi^l,\qquad W^{kl}=W^k-W^l \eqno(4.10)
$$
Eqs (4.6, 4.9, 4.10) define the non-holomorphic effective potential of
N=4 SYM theory in terms of N=1 projections of N=2 superfield strengths.

The last step is restoration of N=2 form of effective action (4.9). For
this purpose we write contribution of non-holomorphic effective potential
to effective action in terms of covariantly constant N=1 projections
$\Phi$ and, $W_\alpha$
$$
\int d^4xd^8\theta{\cal H}(\bar{W},W)=\int d^8zW^\alpha W_\alpha
\bar{W}_{\dot{\alpha}}\bar{W}^{\dot{\alpha}}\frac
{\partial^4{\cal
H}(\bar{\Phi},\Phi)}{\partial\Phi^2\partial\bar{\Phi}^2}+
{\rm derivatives} \eqno(4.11)
$$
Comparison of eqs (4.6, 4.9) and (4.11) leads to
$$
\Gamma^{(1)}=\int d^4xd^8\theta{\cal H}(\bar{W},W)
$$
$$
{\cal H}(\bar{W},W)=\frac{1}{(8\pi)^2}\sum\limits_{k<l}
\log\left(\frac{\bar{W}^k-\bar{W}^l}{\Lambda}\right)^2
\log\left(\frac{W^k-W^l}{\Lambda}\right)^2 \eqno(4.12)
$$
Eq (4.12) is our final result.

\begin{center}
{\large\bf 5. Discussion}
\end{center}

Eq. (4.12) defines the non-holomorphic effective potential depending on
N=2 superfield strengths for N=4 SU(n) super Yang-Mills theories. As a
result we answered all the questions formulated in section 3. First, we
have presented the calcualtional procedure allowing to find
non-holomorphic effective potential. Second, we calculated the
coefficient c in eq. (3.5) for SU(2) group. It is equal to
$1/(8\pi)^2$. Third, a structure of non-holomorphic effective potential
for the gauge group SU(n) has been established.

It is interesting to point out that the scale $\Lambda$ is absent when
the non-holomorphic effective potential (4.12) is written in terms of
N=1 projections of $W$ and $\bar W$ (see eqs (4.6, 4.9)). Therefore, the
$\Lambda$ will be also absent if we write the non-holomorphic effective
potential through the components fields. We need in $\Lambda$ only to
present the final result in manifestly N=2 supersymmetric form.

N=1 form of non-holomorphic effective potential (4.6, 4.9) allows very
easy to get leading bosonic component contribution. Schematically it
has the form $F^4/|\phi|^4$, where $F_{mn}$ is abelian strength
constructed from vector component and $\phi$ is a scalar component of
N=2 vector multiplet ( see also ref [22]). It means that non-zero
expectation value of scalar field $\phi$ plays a role of effective
infrared regulator in N=4 SYM theories.

Generalization of low-energy effective action discussed here and in
refs [18-26] has recently been constructed in ref [27].

{\bf Acknowledgements}

I am very
grateful to E.I. Buchbinder and S.M. Kuzenko for collaboration on
problem of non-holomorphic effective potential and to E.A. Ivanov, B.A.
Ovrut and A.A. Tseytlin for valuable discussions. The work was
supported in part by the RFBR grant \symbol{242} 99-02-16617, by the
DFG-RFBR grant \symbol{242} 99-02-04022, by the INTAS grant No 96-0308
and by GRACENAS grant \symbol{242} 97-6.2-34.

\end{document}